\RequirePackage{fix-cm} 
\RequirePackage[reqno]{amsmath}
\documentclass[final, pdftex, a4paper, twoside, reqno, 12pt]{amsart}

\usepackage[T1]{fontenc}
\usepackage[latin1]{inputenc}
\usepackage{lmodern}

\usepackage[activate, final, kerning, spacing, factor = 1100, stretch = 10, shrink = 10]{microtype}
\microtypecontext{spacing=nonfrench}
\SetTracking{encoding = {*}, shape = sc}{60}

\usepackage[nonewpage, xindy]{imakeidx}
\makeindex[title=Alphabetical index, intoc]

\usepackage{esint}
\usepackage{amsbsy}
\usepackage{amsthm}
\usepackage{cancel}
\usepackage{dsfont}
\usepackage{empheq}
\usepackage{manfnt}
\usepackage{pifont}
\usepackage{xspace}
\usepackage{amssymb}
\usepackage{upgreek}
\usepackage{mathrsfs}
\usepackage{stmaryrd}
\usepackage{etoolbox}
\usepackage{extarrows}
\usepackage{textgreek}

\usepackage[nice]{nicefrac}
\usepackage{siunitx}
\usepackage{fourier-orns}

\usepackage{relsize}
\usepackage{textcomp}

\usepackage{expdlist}
\usepackage{etaremune}

\usepackage{mathtools}
\mathtoolsset{centercolon, mathic}

\usepackage[a4paper, margin=1.0in, bottom=1.33in]{geometry}
\usepackage[dvipsnames, table]{xcolor}
\usepackage[most]{tcolorbox}

\definecolor{bckg}{RGB}{20.8, 20.8, 20.8}
\definecolor{oneblue}{rgb}{0.0, 0.0, 0.85}
\definecolor{Lightblue}{RGB}{214, 214, 214}
\definecolor{bluepigment}{rgb}{0.2, 0.2, 0.6}
\definecolor{charcoal}{rgb}{0.21, 0.27, 0.31}
\definecolor{denimblue}{rgb}{0.08, 0.38, 0.74}
\definecolor{darkelectricblue}{rgb}{0.33, 0.41, 0.47}
\definecolor{katyblue}{rgb}{0.129412, 0.137255, 0.63}

\usepackage{booktabs}
\usepackage{array}
\usepackage{ellipsis}
\usepackage{multicol}

\usepackage{paralist}
\usepackage{csquotes}

\definecolor{mymauve}{rgb}{0.58, 0.0, 0.82}
\usepackage{float}
\usepackage{graphicx}
\graphicspath{{figs/}}
\DeclareGraphicsExtensions{.pdf, .eps}

\usepackage{tikz}
\usepackage{tikz-cd}
\usepackage{subfigure}
\usepackage{morefloats}
\usepackage{indentfirst}

\usepackage[labelsep=period,
            font=normalsize,
            labelformat=simple,
            labelfont={bf,sf,color=bluepigment},
            justification=raggedright]{caption}

\usepackage[bottom, perpage, multiple, symbol]{footmisc}
\setfnsymbol{wiley}

\newcommand*{\Title}{\textcolor{bluepigment}{Integrability of an extended BBM-type equation}}
\newcommand*{\Longtitle}{On integrability of a new dynamical system associated with the BBM-type hydrodynamic flow}
\newcommand*{\Authors}{\textcolor{bluepigment}{D.~Dutykh and Ya.~Prykarpatskyy}}
\newcommand*{\plogo}{\textcolor{gray}{{\texttt{arXiv.org} / \textsc{hal}}}}
\newcommand*{\Keywords}{nonlinear dispersive waves; BBM-type equations; integrable models; Lie symmetries analysis; local symmetries; Lie--B\"acklund symmetry; Poisson structure; bi-Hamiltonian systems}

\usepackage[sort&compress, comma, square, numbers]{natbib}
\makeatletter
\renewcommand{\@biblabel}[1]{\textbf{[#1]}}
\makeatother

\usepackage{url}  
\usepackage[final=true,
           linktoc=all,
           unicode=true,
           bookmarks=true,
           colorlinks=true,
           breaklinks=true,
           urlcolor=katyblue,
           linkcolor=MidnightBlue,
           citecolor=NavyBlue,
           pdffitwindow=false,
           bookmarksopen=false,
           pdfpagemode=UseNone,
           plainpages=false,
           hyperfigures=true,
           pdfstartview={FitV},
           pdftitle={\Longtitle},
           pdfauthor={Denys DUTYKH},
           pdfsubject={Applied Mathematics},
           pdfcreator={Dr. D},
           pdfproducer={pdfLaTeX},
           pdfkeywords={\Keywords},
           pdfnewwindow=true,
           backref=page,
           pagebackref=true, pdfa]{hyperref}

\usepackage[sort&compress, capitalise, noabbrev]{cleveref}

\usepackage[all]{hypcap}
\usepackage{doi}

\usepackage[explicit]{titlesec}

\newcommand\invisiblesection[1]{%
  \addcontentsline{toc}{section}{#1}%
  \sectionmark{#1}}

\titleformat{\section}
  {\color{NavyBlue}\Large\sffamily\bfseries}
  {}
  {0em}
  {\hspace*{-0.5em}\colorbox{bckg!5}{\parbox{\dimexpr\linewidth+0\fboxsep\relax}{\centering\Large\thesection. #1}}}
  [\vspace*{0.33em}]

\titleformat{name=\section, numberless}
  {\color{NavyBlue}\Large\sffamily\bfseries}
  {}
  {0.0em}
  {\hspace*{-0.5em}\colorbox{bckg!5}{\parbox{\dimexpr\linewidth+0\fboxsep\relax}{\centering#1}}}
  [\vspace*{0.33em}]

\titleformat{\subsection}
  {\color{NavyBlue}\large\sffamily\bfseries}
  {}
  {0.0em}
  {\hspace*{-0.5em}\colorbox{bckg!5}{\parbox{\dimexpr\linewidth+0\fboxsep\relax}{\centering\thesubsection. #1}}}
  [\vspace*{0.33em}]

\titleformat{name=\subsection, numberless}
  {\color{NavyBlue}\Large\sffamily\bfseries}
  {}
  {0em}
  {\hspace*{-0.5em}\colorbox{bckg!5}{\parbox{\dimexpr\linewidth+0\fboxsep\relax}{\centering#1}}}
  [\vspace*{0.33em}]

\titleformat{\subsubsection}
  {\color{bluepigment}\sffamily\normalsize\bfseries}
  {}
  {0.0em}
  {\hspace*{-0.5em}\colorbox{bckg!5}{\parbox{\dimexpr\linewidth+0\fboxsep\relax}{\centering\thesubsubsection. #1}}}
  [\vspace*{0.33em}]

\titleformat{name=\subsubsection, numberless}
  {\color{bluepigment}\sffamily\normalsize\bfseries}
  {}
  {0.0em}
  {\hspace*{-0.5em}\colorbox{bckg!5}{\parbox{\dimexpr\linewidth+0\fboxsep\relax}{\centering#1}}}
  [\vspace*{0.33em}]

\titleformat{\paragraph}[runin]
  {\color{bluepigment}\sffamily\small\bfseries}
  {}
  {0em}
  {#1}

\titlespacing{\section}{1.0em}{1.5em plus 2pt minus 2pt}%
{1.0em plus 2pt minus 2pt}[0em]
\titlespacing{\subsection}{1.0em}{1.5em plus 2pt minus 2pt}%
{1.0em plus 2pt minus 2pt}[0em]
\titlespacing{\subsubsection}{1.0em}{1.5em plus 2pt minus 2pt}%
{1.0em plus 2pt minus 2pt}[0em]

\usepackage{titletoc}

\contentsmargin{0.5em}
\newlength{\tocsep} 
\titlecontents{section}[\tocsep]
  {\addvspace{10pt}\bfseries\sffamily}
  {\contentslabel[\thecontentslabel]{\tocsep}}
  {}
  {\ \titlerule*[0.75pc]{.}\ \thecontentspage}
  []

\titlecontents{subsection}[\tocsep]
  {\addvspace{8pt}\sffamily}
  {\contentslabel[\thecontentslabel]{\tocsep}}
  {}
  {\ \titlerule*[0.5pc]{.}\ \thecontentspage}
  []

\titlecontents{subsubsection}[\tocsep]
  {\addvspace{4pt}\footnotesize\sffamily}
  {\contentslabel[\thecontentslabel]{\tocsep}\hspace*{0.5em}}
  {}
  {\ \titlerule*[0.35pc]{.}\ \thecontentspage}
  []

\makeatletter
\def\@setauthors{%
  \begingroup
  \def\thanks{\protect\thanks@warning}%
  \trivlist
  \centering\footnotesize \@topsep30\p@\relax
  \advance\@topsep by -\baselineskip
  \item\relax
  \author@andify\authors
  \def\\{\protect\linebreak}%
  \textsc{\normalsize\textcolor{charcoal}{\authors}}%
  \ifx\@empty\contribs
  \else
    ,\penalty-3 \space \@setcontribs
    \@closetoccontribs
  \fi
  \endtrivlist
  \endgroup
}
\def\@settitle{\begin{center}%
  \baselineskip14\p@\relax
    \bfseries
    \textsc{\Large\textcolor{charcoal}{\@title}}
  \end{center}%
}
\makeatother

\usepackage{enumitem}
\setlist[description]{%
  topsep = 9pt,               
  itemsep = 7pt,               
  labelsep = 10pt,
  font={\bfseries\color{NavyBlue}}, 
}
\setlist[itemize]{%
  itemsep = 5pt,
  font={\color{NavyBlue}}
}

\usepackage{fancyhdr}
\usepackage{lastpage}

\usepackage{xpatch}
\xpretocmd\headrule{\color{lightgray}}{}{\PatchFailed}

\numberwithin{equation}{section}

\theoremstyle{plain}
\newtheorem{theorem}{Theorem}[section]

\newtheorem{prop}{Proposition}[section]
\newtheorem{corollary}{Corollary}[section]

\theoremstyle{definition}

\theoremstyle{remark}

\usepackage{setspace}
\vfuzz \hfuzz
\newcommand{\nm}[1]{\textsc{#1}}

\newcommand{\up}[1]{$^{\mathrm{\small\textsf{\,#1}}}$} 

\newtcbox{\mymath}[1][]{%
    nobeforeafter, math upper, tcbox raise base,
    enhanced, colframe = black!35,
    colback = black!5, boxrule = 1pt, arc = 0mm,
    #1}

\newcommand{\C}{\mathds{C}}
\newcommand{\M}{\mathds{M}}
\newcommand{\T}{\mathds{T}}
\newcommand{\R}{\mathds{R}}
\newcommand{\Z}{\mathds{Z}}
\newcommand{\ud}{\mathrm{d}}
\newcommand{\F}{\mathcal{F}}
\newcommand{\K}{\mathscr{K}}
\newcommand{\Q}{\mathcal{Q}}
\newcommand{\Kk}{\mathcal{K}}
\renewcommand{\H}{\mathcal{H}}
\renewcommand{\O}{\mathcal{O}}
\newcommand{\fO}{\underline{0}}
\newcommand{\Id}{\mathds{1}}

\renewcommand{\geq}{\geqslant}

\newcommand{\cf}{\emph{cf.}\xspace}
\newcommand{\ie}{\emph{i.e.}\xspace}
\newcommand{\eg}{\emph{e.g.}\xspace}
\newcommand{\etc}{\emph{etc.}\xspace}

\renewcommand{\sim}{\thicksim}

\newcommand{\sech}{\mathrm{sech}}

\newcommand{\dprime}{\prime\prime}

\newcommand{\grad}{\boldsymbol{\nabla}}

\newcommand{\defeq}{\mathop{\stackrel{\,\mathrm{def}}{\eqcolon}\,}}
\newcommand{\eqdef}{\mathop{\stackrel{\,\mathrm{def}}{\coloneq}\,}}
\newcommand{\pd}[2]{\frac{\partial\hspace{0.0556em} #1}{\partial\/ #2}}

\newcommand{\od}[2]{\frac{\mathrm{d}\hspace{0.0556em} #1}{\mathrm{d}\/#2}}

\DeclareFontFamily{U}{MnSymbolC}{}
\DeclareSymbolFont{MnSyC}{U}{MnSymbolC}{m}{n}
\DeclareFontShape{U}{MnSymbolC}{m}{n}{
    <-6>  MnSymbolC5
   <6-7>  MnSymbolC6
   <7-8>  MnSymbolC7
   <8-9>  MnSymbolC8
   <9-10> MnSymbolC9
  <10-12> MnSymbolC10
  <12->   MnSymbolC12}{}
\DeclareMathSymbol{\intprod}{\mathbin}{MnSyC}{'270}

\DeclarePairedDelimiterX\abs[1]\lvert\rvert{
  \ifblank{#1}{\:\cdot\:}{\,#1\,}
}
\DeclarePairedDelimiterX\norm[1]\lVert\rVert{
  \ifblank{#1}{\:\cdot\:}{\,#1\,}
}
\DeclarePairedDelimiterX\set[1]{\lbrace}{\rbrace}{\,#1\,}
\DeclarePairedDelimiterX\floor[1]{\lfloor}{\rfloor}{\,#1\,}
\DeclarePairedDelimiterX\Inner[2]{\langle}{\rangle}{\,#1,\,#2\,}
\DeclarePairedDelimiterX\Set[2]{\lbrace}{\rbrace}{\,#1\ \delimsize\vert\ #2\:}
\DeclarePairedDelimiterX\lb[2]{[}{]}{\,\ifblank{#1}{-}{#1}\,,\,\ifblank{#2}{-}{#2}\,}
\DeclarePairedDelimiterX\pb[2]{\lbrace}{\rbrace}{\,\ifblank{#1}{-}{#1}\,,\,\ifblank{#2}{-}{#2}\,}

\makeatletter
\newcommand{\etabar}{\text{\eta@bar}}
\newcommand{\eta@bar}{%
  \vphantom{$\m@th \eta$}%
  \ooalign{%
    $\m@th \eta$\cr
    \hidewidth\kern.25em\smash{\raisebox{-0.7ex}{$\m@th\mathchar'55$}}\hidewidth\cr}%
}
\makeatother

\newcommand{\half}{{\textstyle{\frac{1}{2}}}}
\newcommand{\third}{{\textstyle{\frac{1}{3}}}}
\newcommand{\twothirds}{{\textstyle{\frac{2}{3}}}}

\makeatletter
\renewcommand*\env@matrix[1][\arraystretch]{%
  \edef\arraystretch{#1}%
  \hskip -\arraycolsep
  \let\@ifnextchar\new@ifnextchar
  \array{*\c@MaxMatrixCols c}}
\makeatother

\makeatletter
\renewenvironment{abstract}{%
    \small\thispagestyle{empty}
    \null\vfil
    {\textcolor{RoyalBlue}{\scshape\abstractname.}}
    \quotation
    }
{\endquotation\vfil\null\clearpage}
\makeatother

\newcommand{\TheEnd}{
\bigskip\bigskip
\begin{center}
  \Large
  \decofourleft\hspace*{0.5em}\floweroneleft\hspace*{0.5em}\decoone\hspace*{0.5em}\floweroneright\hspace*{0.5em}\decofourright
\end{center}
\bigskip\bigskip}

\begin{document}

\title[\Title]{\Longtitle}

\author[D.~Dutykh]{Denys Dutykh\textcolor{denimblue}{$^*$}}
\address{\textcolor{denimblue}{\bfseries D.~Dutykh:} Mathematics Department, Khalifa University of Science and Technology, PO Box 127788, Abu Dhabi, United Arab Emirates \and Causal Dynamics Pty Ltd, Perth, Australia}
\email{\href{mailto:Denys.Dutykh@ku.ac.ae}{Denys.Dutykh@ku.ac.ae}}
\urladdr{\url{https://www.denys-dutykh.com/}}
\thanks{\textcolor{denimblue}{$^*$}\itshape Corresponding author}

\author[Ya.~A.~Prykarpatskyy]{Yarema A. Prykarpatskyy}
\address{\textcolor{denimblue}{\bfseries Ya.~A.~Prykarpatskyy:} The Department of Applied Mathematics, University of Agriculture in Krak\'{o}w, al. Mickiewicza 21, 31-120, Krak\'{o}w, Poland \&
Institute of Mathematics of NAS of Ukraine, Tereschenkivska 3, Kiev-4,
01024, Ukraine
}
\email{\href{mailto:yarema.prykarpatskyy@urk.edu.pl}{Yarema.Prykarpatskyy@urk.edu.pl}}
\urladdr{\url{https://kzmmoodle.dynu.net/~yarpry/}}



\begin{titlepage}
\clearpage
\pagenumbering{arabic}
\thispagestyle{empty} 
\noindent
{\Large Denys \textsc{Dutykh}}
\\[0.001\textheight]
{\textit{\textcolor{gray}{Khalifa University of Science and Technology, Abu Dhabi, UAE}}}
\\[0.02\textheight]
{\Large Yarema A. \nm{Prykarpatskyy}}
\\[0.001\textheight]
{\textit{\textcolor{gray}{University of Agriculture in Krak\'{o}w, Krak\'{o}w, Poland}}}
\\[0.16\textheight]

\vspace*{0.16cm}

\colorbox{Lightblue}{
  \parbox[t]{1.0\textwidth}{
    \centering\huge
    \vspace*{0.75cm}
    
    \textsc{\textcolor{katyblue}{\Longtitle}}
    
    \vspace*{0.75cm}
  }
}

\vfill 

\raggedleft     
{\large \plogo} 
\end{titlepage}


\clearpage
\thispagestyle{empty} 
\par\vspace*{\fill}   
\begin{flushright} 
{\textcolor{RoyalBlue}{\textsc{Last modified:}} \today}
\vspace*{1.0em}
\end{flushright}


\clearpage
\maketitle
\thispagestyle{empty}


\begin{abstract}

This article explores the exceptional integrability property of a family of higher-order \nm{Benjamin--Bona--Mahony} (BBM)-type nonlinear dispersive equations. Here, we highlight its deep relationship with a generalized infinite hierarchy of the integrable \nm{Riemann}-type hydrodynamic equations. A previous \nm{Lie} symmetry analysis revealed a particular case which was conjectured to be integrable. Namely, a \nm{Lie--B\"acklund} symmetry exists, thus highlighting another associated dynamical system. Here, we investigate these two equations using the gradient-holonomic integrability scheme. Moreover, we construct their infinite hierarchy of conservation laws analytically, using three compatible \nm{Poisson} structures to prove the complete integrability of both dynamical systems. We investigate these two equations using the so-called gradient-holonomic integrability scheme. Based on this scheme, applied to the equation under consideration, we have analytically constructed its infinite hierarchy of conservation laws, derived two compatible \nm{Poisson} structures and proved its complete integrability.


\bigskip\bigskip
\noindent \textcolor{RoyalBlue}{\textbf{\keywordsname:}} \Keywords \\

\smallskip
\noindent \textcolor{RoyalBlue}{\textbf{MSC:}} 
\smallskip \\
\noindent \textcolor{RoyalBlue}{\textbf{PACS:}} 

\end{abstract}


\section{Introduction}

The water wave theory has always been developed by constructing more and more peculiar approximations \cite{Craik2004}. Among the small amplitude (\ie weakly nonlinear) and weakly dispersive unidirectional models, the two model equations share the throne: the celebrated \nm{Korteweg--de Vries} \cite{KdV, Boussinesq1877}
\begin{equation*}
  u_{\,t}\ +\ u_{\,x}\ +\ u\,u_{\,x}\ +\ u_{\,x\,x\,x}\ =\ \fO
\end{equation*}
and \nm{Benjamin--Bona--Mahony} \cite{Peregrine1966, bona}
\begin{equation}\label{eq:bbm}
  u_{\,t}\ +\ u_{\,x}\ +\ u\,u_{\,x}\ -\ u_{\,t\,x\,x}\ =\ \fO
\end{equation}
equations. Here, the variable $u$ may refer to the horizontal (\eg depth-averaged) velocity or free surface elevation. The first one is integrable, possesses an infinite hierarchy of (local) conservation laws and \nm{Galilean} invariant. The second equation has none of these properties. Regarding the integrability (and the infinite hierarchy of conservation laws as a consequence), we all agree that it is an exceptional property. However, the \nm{Galilean} invariance property must be present in any physically sound model. It seems to be completely normal that two observers (one sitting on a chair and another walking along a channel with constant speed) see the same waves (and flow) modulo a constant shift in the velocity. In other words, the laws of (classical) Physics are identical in all inertial frames of reference. To make a long story short, the \nm{Galilean}-invariant (but higher order) iBBM equation was proposed in \cite{Duran2013}. However, this equation had at least one shortcoming --- it did not possess the energy conservation property. The quest for physically sound generalizations was pursued in \cite{Cheviakov2023} with some partial results announced recently in \cite{Cheviakov2021}. Namely, the following family of Partial Differential Equations (PDEs) has been proposed and studied in \cite{Cheviakov2023, Cheviakov2021}:
\begin{equation}\label{eq:abbm}
  u_{\,t}\ -\ u_{\,x\,x\,t}\ =\ -\,u_{\,x}\ -\ u\,u_{\,x}\ +\ \alpha\,(\,u\,u_{\,x\,x\,x}\ +\ 2\,u_{\,x}\,u_{\,x\,x}\,)\,,
\end{equation}
where $\alpha\ \in\ \R$ is a real parameter, $u\ \in\ C^{\infty}\,(\,\R\,;\,\R\,)$ is a smooth real valued function. Symbols $u_{\,t}\,$, $u_{\,x}\,$, $u_{\,x\,x}\,$, \etc denote the partial derivatives of $u$ with respect to the independent spatial $x\ \in\ \R$ and temporal $t\ \in\ \R$ variables. The symmetry and conservation laws analysis has shown that there are two special cases with respect to the parameter $\alpha$ \cite{Cheviakov2021}:
\begin{description}
  \item[$\alpha\ =\ 1\,$] The eBBM equation admitting the \nm{Galilean} invariance and energy conservation. This case will be studied in \cite{Cheviakov2023}:
  \begin{equation}\label{eq:ebbm}
    u_{\,t}\ -\ u_{\,x\,x\,t}\ =\ -\,u_{\,x}\ -\ u\,u_{\,x}\ +\ u\,u_{\,x\,x\,x}\ +\ 2\,u_{\,x}\,u_{\,x\,x}\,.
  \end{equation}
  \item[$\alpha\ =\ \frac{1}{3}\,$] This case is represented by the following equation:
  \begin{equation}\label{eq:ebbm-0}
    u_{\,t}\ -\ u_{\,x\,x\,t}\ =\ -\,u_{\,x}\ -\ u\,u_{\,x}\ +\ \frac{1}{3}\;u\,u_{\,x\,x\,x}\ +\ \frac{2}{3}\;u_{\,x}\,u_{\,x\,x}\,.
  \end{equation}
  It was shown in \cite{Cheviakov2021} that this equation admits an additional
\nm{Lie--B\"acklund} symmetry. It is suspected to be integrable and will be
discussed below.
\end{description}
Of course, the classical BBM equation is recovered for the parameter $\alpha\ =\ 0\,$.

Having done the change of the temporal variable $t\ \rightleftharpoons\ 3\,\tau
\in \R\,$, the Equation~\eqref{eq:ebbm-0} can be rewritten as
\begin{equation*}
  u_{\,\tau }\ -\ u_{\,x\,x\,\tau }\ =\ -\,u_{\,x}\ -\ 3u\,u_{\,x}\ +\ u\,u_{\,x\,x\,x}\ +\ 2\;u_{\,x}\,u_{\,x\,x}
\end{equation*}
or equivalently as the equation
\begin{equation}\label{eq-2}
  D_{\,\tau}\,z\ =\ -\,2\,u_{\,x}\,z\,,
\end{equation}
where we introduced the following notation
\begin{equation*}
  z\ \eqdef\ u\ -\ u_{\,x\,x}\ +\ \half\,, \qquad
  D_{\,\tau}\ \eqdef\ \partial_{\,\tau}\ +\ u\,\partial_{\,x}\,.
\end{equation*}
The operator $D_{\,\tau}$ is nothing else but the usual convection operator. The obtained differential expression \eqref{eq-2} can be naturally extended via the evolution relation $D_{\,\tau}\,u\ =\ z$ on the function $u\ \in \ C^{\,\infty}\,(\,\R\,;\,\R\,)$ to the following system:
\begin{equation}\label{eq-3}
  D_{\,\tau}\,u\ =\ z\,, \qquad 
  D_{\,\tau}\,z\ =\ -\,2\,u_{\,x}\,z\,.
\end{equation}
The last evolution system was obtained and discussed in detail some years ago in \cite{Blackmore2014, Blackmore2011} as a special case of the following integrable generalized \nm{Riemann}-type hydrodynamic system:
\begin{equation}\label{eq:0.8}
  D_{\,\tau}^{\,N-1}\,u\ =\ \bar{z}_{x}^{\,s}\,, \qquad
  D_{\,\tau}\,\bar{z}\ =\ \fO\,,
\end{equation}
for $N\ =\ 2\,$, $s\ =\ 2$ and
\begin{equation*}
  \bar{z}_{\,x}^{\,s}\ \eqdef\ =\ u\ -\ u_{\,x\,x}\,.
\end{equation*}
Moreover, there was obtained the \nm{Lax}-type representation of \eqref{eq-3}
\begin{equation*}
  \left[\,D_{\,x}\ -\ \left(
\begin{array}{cc}
\bar{z}_{\,x} & \fO \\
-\lambda (u+u_{\,x}/\bar{z}_{\,x}) & -\bar{z}_{\,x\,x}/\bar{z}_{\,x}%
\end{array}\right),\;
D_{\,\tau}\ -\ \left(
\begin{array}{cc}
0 & 0 \\
-\lambda\,\bar{z}_{\,x} & u_{\,x}
\end{array}
\right)\,\right]\ =\ \fO\,,
\end{equation*}
compatible for all $\lambda\ \in\ \R\,$. The result above strictly means that the integrability of the extended eBBM-equation \eqref{eq:ebbm-0} is more than argued and what will be demonstrated directly using the gradient-holonomic scheme, proposed in \cite{Mitropolsky1987, Prykarpatsky1998, Blackmore2011}.

It is worth mentioning that similar results we obtained in \cite{Blackmore2014} for the celebrated integrable \nm{Degasperis}--\nm{Procesi} equation \cite{Degasperis1999} along with its infinite hierarchy generalizations. For instance, take Equations~\eqref{eq:0.8} with $N\ =\ 3$ and $s\ =\ 3\,$:
\begin{equation*}
  D_{\,\tau}\,u\ =\ \bar{z}_{x}^{\,3}\,, \qquad
  D_{\,\tau}\,\bar{z}\ =\ \fO
\end{equation*}
and put $z\ \eqdef\ \bar{z}_{x}^{\,3}\,$. The resulting system will then read
\begin{equation*}
  D_{\,\tau}\,u\ =\ z\,, \qquad
  D_{\,\tau}\,\bar{z}\ =\ -\,3\,u_{\,x}\,z\,.
\end{equation*}
If now we make the reduction $z\ \eqdef\ u\ -\ u_{\,x\,x}\,$, the last equation transforms into the following scalar PDE:
\begin{equation*}
  u_{\,t}\ -\ u_{\,x\,x\,t}\ =\ -\,4\,u\,u_{\,x}\ +\ u\,u_{\,x\,x\,x}\ +\ 3\,u_{\,x}\,u_{\,x\,x}\,.
\end{equation*}
The last equation is nothing else but the famous \nm{Degasperis}--\nm{Procesi} equation \cite{Degasperis1999}.

Another interesting illustration follows from the generalized integrable \nm{Riemann}-type Equation~\eqref{eq:0.8} written in a slightly modified form:
\begin{equation}\label{eq:0.13}
  D_{\,\tau}^{\,N-1}\,v\ =\ \bar{z}_{x}^{\,s}\,, \qquad
  D_{\,\tau}\,\bar{z}\ =\ \fO\,,
\end{equation}
with $D_{\,\tau}\ \eqdef\ \partial_{\,t}\ +\ v\,\partial_{\,x}\,$. If we specify $N\ =\ 2$ and $s\ =\ 1\,$, we obtain:
\begin{equation*}
  D_{\,\tau}\,v\ =\ \bar{z}_{x}\,, \qquad
  D_{\,\tau}\,\bar{z}\ =\ \fO\,.
\end{equation*}
After the substitution $\bar{z}_{\,x}\ \leftarrow\ z$ and $v_{\,x}\ \leftarrow\ 2\,u_{\,x}\,z\,$, one obtains the following interesting integrable system:
\begin{equation*}
  D_{\,\tau}\,v\ =\ z\,, \qquad
  z_{\,t}\ +\ 2\,[\,z\,\partial^{\,-\,1}\,(\,z\,u_{\,x}\,)\,]_{\,x}\ =\ \fO\,.
\end{equation*}
What is surprising is that the next reduction $z\ \eqdef\ u\ -\ u_{\,x\,x}$ gives rise to the following integrable equation:
\begin{equation*}
  u_{\,t}\ -\ u_{\,x\,x\,t}\ +\ \Bigl[\,(\,u\ -\ u_{\,x\,x}\,)\,\bigl(\,u^{\,2}\ -\ u_{x}^{\,2}\,\bigr)\,\Bigr]_{\,x}\ =\ \fO\,,
\end{equation*}
whose integrability was established some years ago in \cite{Qiao2006}. Its \nm{Lax}-type representation
\begin{multline*}
  \biggl[\,\partial_{\,x}\ -\ \begin{pmatrix}
    -\,\frac{1}{2} & \frac{\lambda\,z}{2} \\
    -\,\frac{\lambda\,z}{2} & \frac{1}{2}
  \end{pmatrix}\,, \\ 
  \partial_{\,t}\ -\ \begin{pmatrix}
    \lambda^{\,-\,2}\ +\ \partial^{\,-\,1}\,(\,z\,u_{\,x}\,) & -\,\lambda^{\,-\,1}\,(\,u\ -\ u_{\,x}\,)\ -\ z\,\partial^{\,-\,1}\,(\,z\,u_{\,x}\,) \\
    \lambda^{\,-\,1}\,(\,u\ +\ u_{\,x}\,)\ +\ z\,\partial^{\,-\,1}\,(\,z\,u_{\,x}\,) & -\,\lambda^{\,-\,2}\ -\ \partial^{\,-\,1}\,(\,z\,u_{\,x}\,)
  \end{pmatrix}\,\biggr]\ =\ \fO\,.
\end{multline*}
is satisfied for $\forall\,\lambda\ \in\ \R^{\,\times}\,$.

Some years ago, a ``new'' generalized integrable \nm{Camassa}--\nm{Holm}-type equation was announced in \cite{Novikov2009}:
\begin{equation}\label{eq:0.18}
  u_{\,t}\ -\ u_{\,x\,x\,t}\ =\ u^{\,2}\,u_{\,x\,x\,x}\ +\ 3\,u\,u_{\,x}\,u_{\,x\,x}\ -\ 4\,u^{\,2}\,u_{\,x}\,.
\end{equation}
The last equation can also be obtained as a reduction of the generalized integrable \nm{Riemann}-type hydrodynamic Equation~\eqref{eq:0.13} with $N\ =\ 2$ and $s\ =\ -\,2$ on the functional sub-manifold $z\ \eqdef\ \bar{z}_{x}^{\,-\,2}\,$:
\begin{equation*}
  D_{\,\tau}\,v\ =\ z\,, \qquad
  D_{\,\tau}\,z\ =\ 2\,v_{\,x}\,z\,.
\end{equation*}
After the change of variables $v\ \leftarrow\ u^{\,2}$ and $z\ \leftarrow\ u\ -\ u_{\,x\,x}$ we obtain \cref{eq:0.18}.

The present manuscript is organized as follows. The main relevant properties of governing equations are outlined in \cref{sec:eqs}. The integrability of two dynamical systems under consideration is proved in \cref{sec:int}. The integrability property is highlighted on the solitary wave collisions in \cref{sec:num}, and solitary wave solutions are studied analytically and numerically. Finally, this study's main conclusions and perspectives of this study are outlined in \cref{sec:concl}.


\section{Governing equations}
\label{sec:eqs}

From now on, we consider a particular representative of the $\alpha-$family of PDEs \eqref{eq:abbm} with $\alpha\ =\ \frac{1}{3}\,$, which was spotted thanks to the computer-aided \nm{Lie} point symmetry analysis \cite{Cheviakov2021}:
\begin{equation}\label{eq:bbm3}
  u_{\,t}\ -\ u_{\,x\,x\,t}\ =\ -\,u_{\,x}\ -\ u\,u_{\,x}\ +\ \frac{1}{3}\;u\,u_{\,x\,x\,x}\ +\ \frac{2}{3}\;u_{\,x}\,u_{\,x\,x}\,.
\end{equation}
In previous works \cite{Cheviakov2023, Cheviakov2021} this equation was referred to as the eBBM${}_{\,\third}$ equation. It possesses both a wide hierarchy of \nm{Lie} symmetries and conservation laws, amongst which
\begin{align*}
  \tilde{\gamma}_{\,1}\ &\coloneq\ \int_{\,T}\,\bigl[\,5\,u^{\,4}\ +\ 12\,u^{\,3}\ +\ u^{\,2}\,\bigl(\,26\,u_{x}^{\,2}\ +\ 4\,u_{x\,x}^{\,2}\,\bigr)\,u^{\,2}\ +\\ 
  & \qquad\qquad u\,\bigl(\,u_{x}^{\,2}\,u_{\,x\,x}\ +\ 24\,u_{x}^{\,2}\ +\ 36\,u_{\,t\,t}\,u_{\,x\,x}\,\bigr)\ -\ 36\,u_{\,t\,t}\,u_{\,x\,x}\,\bigr]\,\ud x\,, \\
  \tilde{\gamma}_{\,2}\ &\coloneq\ \int_{\,T}\,\sqrt{\,2\,(\,u\ -\ u_{\,x\,x}\,)\ +\ 3}\;\ud x\,, \\
  \tilde{\gamma}_{\,3}\ &\coloneq\ \int_{\,T}\,\bigl(\,2\,(\,u\ -\ u_{\,x\,x}\,)\ +\ 3\,\bigr)^{\,\frac{5}{2}}\,\bigl[\,(2\,(\,u\ -\ u_{\,x\,x}\,)\ +\ 3)^{\,2}\ + \\ 
  & \qquad\qquad (\,2\,u\ +\ 3\,)\,(\,2\,(\,u\ -\ u_{\,x\,x}\,)\ +\ 3\,)\ +\ 3\,(\,u_{x\,x\,x}^{\,2}\ -\ u_{x}^{\,2}\,)\,\bigr]\,\ud x\,,
\end{align*}
where the integration domain $T$ is the whole real line $\R$ or a 1D periodic torus. In the first case, we have to add the corresponding assumptions on the decaying behaviour of all the functions at infinity. 

Moreover, \cref{eq:bbm3} possesses also one new \nm{Lie--B\"acklund} symmetry \cite{Olver1993, Cheviakov2021}:
\begin{equation}\label{eq:flow}
  u_{\,\tau}\ =\ \frac{u_{\,x}\ -\ u_{\,x\,x\,x}}{\bigl[\,2\,(\,u\ -\ u_{\,x\,x}\,)\ +\ 3\,]^{\,\frac{3}{2}}}
\end{equation}
with respect to an evolution parameter $\tau\ \in\ \R\,$. Moreover, it was shown that the flow \eqref{eq:abbm} is \nm{Hamiltonian} \cite{Cheviakov2021}:
\begin{equation*}
  u_{\,t}\ =\ -\,\vartheta\,\grad_{\,u}\,\tilde{\H}_{\,1}\,[\,u\,]
\end{equation*}
with respect to the \nm{Hamiltonian} functional
\begin{equation*}
  \tilde{\H}_{1}\ =\ \frac{1}{2}\;\int_{\,T}\,\bigl[\,u^{\,2}\ +\ \third\;\bigl(\,u^{\,3}\ +\ u\,u_{x}^{\,2}\,\bigr)\,\bigr]\,\ud x
\end{equation*}
and the \nm{Poisson} operator
\begin{equation}\label{eq:1.6}
  \vartheta\ \eqdef\ \bigl(\,\partial^{\,-1}\ -\ \partial\,\bigr)^{\,-1}\,,
\end{equation}
where, by definition, $\partial\ \eqdef\ \pd{}{x}\,$, $x\ \in\ \R$ and $\grad_{\,u}$ denotes the variational (\nm{Fr\'echet}) derivative with respect to the function $u\ \in\ C^{\,\infty}\,(\,\R\,;\,\R\,)\,$. For a smooth function $u\ \in\ C^{\,\infty}\,(\,\M\,)\,$, by $[\,u\,]$ we denote its infinite jet. It has been conjectured and confirmed in \cite{Cheviakov2021} that the flow \eqref{eq:bbm3} is closely related to the well-known \nm{Camassa--Holm} equation \cite{Johnson2002, Dullin2003} and is a completely integrable \nm{Hamiltonian} system \cite{Faddeev1987, Novikov1984, Moser2005}.

In the following Section, we are building on the thorough \nm{Lie} symmetry analysis performed in \cite{Cheviakov2021}, especially on the fact that quantities $\tilde{\gamma}_{\,1,\,2,\,3}$ are conserved for both evolution flows \eqref{eq:bbm3} and \eqref{eq:flow}, we provide an analytical proof of complete integrability for the \nm{Hamiltonian} system \eqref{eq:bbm3} and its symmetry flow \eqref{eq:flow}.


\section{Integrability analysis}
\label{sec:int}

First, we observe that the dynamical system \eqref{eq:bbm3} is equivalent to the following nonlinear dynamical system:
\begin{equation}\label{eq:2.1}
  u_{\,t}\ =\ -\,(\,1\ -\ \partial_{x}^{\,2}\,)^{\,-1}\,\bigl[\,u_{\,x}\ +\ u\,u_{\,x}\ -\ \third\;(\,u\,u_{\,x\,x\,x}\ +\ 2\,u_{\,x}\,u_{\,x\,x}\,)\,\bigr]\ \defeq\ \tilde{\F}\,[\,u\,]
\end{equation}
on a functional manifold $\M\ \subseteq\ C^{\,\infty}\,(\,\R\,;\,\R\,)$ of smooth real-valued functions. The last equation possesses as its symmetry the evolution flow
\begin{equation}\label{eq:2.2}
  u_{\,\tau}\ =\ \frac{u_{\,x}\ -\ u_{\,x\,x\,x}}{[\,2\,(\,u\ -\ u_{\,x\,x}\,)\ +\ 3\,]^{\,\frac{3}{2}}}\ \defeq\ \tilde{\Kk}\,[\,u\,]\,,
\end{equation}
which, jointly with \eqref{eq:2.1}, are commuting to each other, \ie
\begin{equation*}
  \lb*{\tilde{\Kk}}{\tilde{\F}}\,[\,u\,]\ \eqdef\ \tilde{\F}^{\,\prime}\,\tilde{\Kk}\,[\,u\,]\ -\ \tilde{\Kk}^{\,\prime}\,\tilde{\F}\,[\,u\,]\ =\ \fO\,,
\end{equation*}
for all smooth functions $u\ \in\ \M\,$. Above, the $(\,-\,)^{\,\prime}$ is a short-hand notation for the \nm{Fr\'echet} derivative of a functional $(\,-\,)\,$.

Moreover, it is easy to check that the flows \eqref{eq:2.1} and \eqref{eq:2.2} above reduce via the argument transformation
\begin{equation*}
  u\,(\,x,\,t\,)\ \rightarrow\ v\,\Bigl(\,x\ -\ \frac{t}{2},\,t\,\Bigr)\ -\ \frac{3}{2}\,, \qquad \forall\,(\,x,\,t\,)\ \in\ \R\times\R
\end{equation*}
to the following equivalent \nm{Hamiltonian} forms:
\begin{subequations}\label{eq:2.4}
\begin{align}
  v_{\,t}\ &=\ -\,(\,\partial^{\,-1}\ -\ \partial\,)^{\,-1}\,\grad_{\,v}\,\H_{\,1}\,[\,v\,]\ =\ -\,\vartheta\,\H_{\,1}\,[\,v\,]\ \defeq\ \F\,[\,v\,]\,, \label{eq:2.4a} \\
  \H_{\,1}\ &\eqdef\ \frac{1}{2}\;\int_{\,T}\,\bigl(\,v^{\,2}\ +\ v_{x}^{\,2}\,\bigr)\,\Bigl(\,\frac{v}{3}\ -\ 1\,\Bigr)\,\ud x\, \label{eq:2.4b}
\end{align}
\end{subequations}
with respect to the evolution parameter $t\ \in\ \R$ and
\begin{subequations}\label{eq:2.5}
\begin{align}
  v_{\,\tau}\ &=\ \partial\,\Bigl(\,\frac{1}{\sqrt{\,v\ -\ v_{\,x\,x}}}\,\Bigr)\ \ \stackrel{(*)}{=}\ -\,\vartheta\,\grad_{\,v}\,\H_{\,2}\,[\,v\,]\ \defeq\ \Kk\,[\,v\,]\,, \label{eq:2.5a} \\
  \H_{\,2}\ &\eqdef\ 2\,\int_{\,T}\,\sqrt{\,v\ -\ v_{\,x\,x}}\:\ud x\,, \label{eq:2.5b}
\end{align}
\end{subequations}
with respect to the evolutionary parameter\footnote{
Let us demonstrate, for example, the equality (*) in \eqref{eq:2.5a}. The \nm{Hamiltonian} $\H_{\,2}$ is defined in \eqref{eq:2.5b}. Let us compute its \nm{Fr\'echet} derivative with respect to the evolution variable $v\,$:
\begin{multline*}
  \grad_{\,v}\,\H_{\,2}\ =\ 2\cdot\bigl(\,\sqrt{\,v\ -\ v_{\,x\,x}}\,\bigr)^{\,\prime\,\ast}\cdot\Id\ =\ 2\cdot\Bigl(\,-\,\frac{1}{2}\,\Bigr)\,\Bigl[\,\frac{1}{\sqrt{\,v\ -\ v_{\,x\,x}}}\;(\,\Id\ -\ \partial^{\,2}\,)\,\Bigr]\cdot\Id\ =\\ 
  -\,(\,\Id\ -\ \partial^{\,2}\,)\cdot\frac{1}{\sqrt{\,v\ -\ v_{\,x\,x}}}\ =\ -\,(\,\partial^{\,-\,1}\ -\ \partial\,)\,\partial\,\Bigl(\,\frac{1}{\sqrt{\,v\ -\ v_{\,x\,x}}}\,\Bigr)\ \equiv\ -\,\vartheta^{\,-\,1}\,\partial\,\Bigl(\,\frac{1}{\sqrt{\,v\ -\ v_{\,x\,x}}}\,\Bigr)\,,
\end{multline*}
where $\Id$ denotes the identity operator. Now, we can readily check the equalities in \eqref{eq:2.5a}:
\begin{equation*}
  v_{\,\tau}\ =\ -\,\vartheta\,\grad_{\,v}\,\H_{\,2}\ =\ -\,\vartheta\,(\,-\,\vartheta^{\,-\,1}\,)\,\partial\,\Bigl(\,\frac{1}{\sqrt{\,v\ -\ v_{\,x\,x}}}\,\Bigr)\,\ \equiv\ \partial\,\Bigl(\,\frac{1}{\sqrt{\,v\ -\ v_{\,x\,x}}}\,\Bigr)\,.
\end{equation*}
This finishes the verification of \cref{eq:2.5}.} $\tau\ \in\ \R\,$, correspondingly. The skew-symmetric \nm{Poisson} operator $\vartheta\,:\ \T^{\,\ast}\,(\,\M\,)\ \longrightarrow\ \T\,(\,\M\,)$ naturally acts as a pseudo-differential expression from the cotangent bundle $\T^{\,\ast}\,(\,\M\,)$ to the tangent bundle $\T\,(\,\M\,)$ over the functional \nm{Poisson} manifold $\M$ and determines the associated \nm{Poisson} bracket:
\begin{equation*}
  \pb{-_{\,1}}{-_{\,2}}\ \eqdef\ \bigl(\,\grad\,(\,-_{\,1}\,),\,\vartheta\,\grad\,(\,-_{\,2}\,)\,\bigr)
\end{equation*}
over the space of functionals on $\M\,$, endowed with the natural bi-linear form (pairing)
\begin{equation}\label{eq:bil}
  \bigl(\,-,\,-\,\bigr)\,:\ \T^{\,\ast}\,(\,\M\,)\times\T\,(\,\M\,)\ \longrightarrow\ \R\,.
\end{equation}

In order to establish the complete integrability of the flows \eqref{eq:2.4} and \eqref{eq:2.5}, we take into account the fact that they commute to each other. Henceforth, it is enough to check the integrability of the second flow \eqref{eq:2.5} (which looks slightly simpler) on the functional manifold $\M$ and to show that the first one \eqref{eq:2.4} enters into an infinite hierarchy of commuting to each other \nm{Hamiltonian} flows generated by \eqref{eq:2.5} \cite{Blackmore2011}.

We are aware of three effective integrability checking schemes \cite{Mitropolsky1987, Dubrovin2006, Shabat1993}. In the present article, we employ the gradient-holonomic integrability verification scheme \cite{Mitropolsky1987}. For this specific verification scheme, one needs to find an asymptotic solution to a \nm{Baker--Akhiezer}-type function $\varphi\,(\,x;\,\lambda\,)\ \in\ \T^{\,\ast}\,(\,\M\,)\,$, $\lambda\ \in\ \Gamma$ \cite{Blackmore2011, Prykarpatsky1998}, defined on the spectrum $\Gamma$ of a suitably chosen \nm{Lax}-type operator \cite{Mitropolsky1987, Novikov1984, Moser2005, Newell1985}, equivalent to the recursion operator \cite{Novikov1984, Newell1985, Faddeev1987}
\begin{equation}\label{eq:lambda}
  \Lambda\,:\ \T^{\,\ast}\,(\,\M\,)\ \longrightarrow\ \T^{\,\ast}\,(\,\M\,)\,.
\end{equation}
This implies that the following functional-operator equations on the manifold $\M$
\begin{equation*}
  \Lambda\,\varphi\,(\,x;\,\lambda\,)\ =\ \xi\,(\,\lambda\,)\,\varphi\,(\,x;\,\lambda\,)
\end{equation*}
and
\begin{equation*}
  \od{\Lambda}{\tau}\ =\ \lb{\Lambda}{\K^{\,\prime\,\ast}}\,,
\end{equation*}
where $\xi\,(\,\lambda\,)\ \in\ \C$ is the corresponding eigenvalue, are compatible. This follows from the determining \nm{Lax--Noether} linear equation\footnote{Here $\K^{\,\prime\,\ast}\,:\ \T^{\,\ast}\,(\,\M\,)\ \longrightarrow\ \T^{\,\ast}\,(\,\M\,)$ is the adjoint cotangent map to the \nm{Fr\'echet} derivative $\K^{\,\prime}\,:\ \T\,(\,\M\,)\ \longrightarrow\ \T\,(\,\M\,)$ defined through the following identity:
\begin{equation*}
  \bigl(\,a\,\vert\,\K^{\,\prime}\,b\,\bigr)\ \equiv\ \bigl(\,\K^{\,\prime\,\ast}\,a\,\vert\,b\,\bigr)\,, \quad \forall\,a\ \in\ \T^{\,\ast}\,(\,\M\,)\,, \quad b\ \in\ \T\,(\,\M\,)\,.
\end{equation*}}
\begin{equation}\label{eq:2.9}
  \varphi_{\,\tau}\ +\ \K^{\,\prime\,\ast}\,\varphi\ =\ \fO
\end{equation}
along with \nm{Magri}-type symmetry hereditary property \cite{Magri1978, Mitropolsky1987, Shabat1993}. The conserved quantities of the dynamical system \eqref{eq:2.5} are recurrently generated by means of the asymptotic as $\Gamma\ \ni\ \lambda\ \to\ \infty$ solution to \eqref{eq:2.9} (\cf \eg \cite{Mitropolsky1987}), representable in our case as
\begin{equation*}
  \varphi\ \sim\ \exp\,\bigl[\,-\,\lambda^{\,3}\,t\ +\ \partial^{\,-\,1}\,(\,\sigma_{\,-\,1}\,\lambda\ +\ \sigma_{\,0}\ +\ \sigma_{\,1}\,\lambda^{\,-\,1}\ +\ \sigma_{\,2}\,\lambda^{\,-\,2}\ +\ \ldots)\,\bigr]\,.
\end{equation*}
The functional
\begin{equation*}
  \gamma_{\,j}\ \eqdef\ \int_{\,T}\,\sigma_{\,j}\,[\,v\,]\,\ud x
\end{equation*}
are the corresponding conservation laws of the \nm{Hamiltonian} system \eqref{eq:2.5}, \ie
\begin{equation*}
  \od{\gamma_{\,j}}{\tau}\biggr\vert_{\,v_{\,\tau}\ =\ \Kk\,[\,v\,]}\ \equiv\ \fO\,, \qquad \forall\,j\ \in\ \Z^{\,+}\,\cup\,\set*{-\,1}\,.
\end{equation*}
In particular, one can easily obtain recurrently that \cite[Section~\textsection\,3.2]{Mitropolsky1987} (see also \cite[Section~\textsection\,5.2.1]{Blackmore2011}):
\begin{align*}
  \sigma_{\,-\,1}\ &=\ 2^{\,\frac{1}{3}}\,\sqrt{\,v\ -\ v_{\,x\,x}}\,, \\
  \sigma_{\,0}\ &=\ \frac{1}{2}\;\od{\ln(\,v_{\,x\,x}\ -\ x\,)}{x}\,, \\
  \sigma_{\,1}\ &=\ \frac{\splitfrac{8\,(\,v\,-\,v_{\,x\,x}\,)\,v_{\,x\,x\,x\,x}\,+\,13\,v_{\,x\,x\,x}^{2}\,-\,26\,v_{\,x}\,v_{\,x\,x\,x}}{\,+\,12\,v_{\,x\,x}^{2}\,-\,16\,v\,v_{\,x\,x}\,+\,13\,v_{\,x}^{2}\,+\,4\,v^{\,2}}}{3\times 2^{\,\frac{10}{3}}\,(\,v\,-\,v_{\,x\,x}\,)^{\,\frac{5}{2}}}\,, \\
  \sigma_{\,2}\ &=\ \ldots
\end{align*}
whence the corresponding conservation laws are given below:
\begin{subequations}\label{eq:2.14}
\begin{align}
  \gamma_{\,-\,1}\ &=\ 2^{\,\frac{1}{3}}\,\int_{\,T}\,\sqrt{\,v\ -\ v_{\,x\,x}}\,\ud x\,=\,2^{\,\frac{1}{3}}\,\H_{\,2}\,, \\
  \gamma_{\,0}\ &=\ \frac{1}{2}\;\int_{\,T}\,\od{\ln(\,v_{\,x\,x}\ -\ x\,)}{x}\,\ud x\,\equiv\,\fO\,, \\
  \gamma_{\,1}\ &=\ \int_{\,T}\,\frac{\splitfrac{8\,(\,v\,-\,v_{\,x\,x}\,)\,v_{\,x\,x\,x\,x}\,+\,13\,v_{\,x\,x\,x}^{2}\,-\,26\,v_{\,x}\,v_{\,x\,x\,x}}{\,+\,12\,v_{\,x\,x}^{2}\,-\,16\,v\,v_{\,x\,x}\,+\,13\,v_{\,x}^{2}\,+\,4\,v^{\,2}}}{3\times 2^{\,\frac{10}{3}}\,(\,v\,-\,v_{\,x\,x}\,)^{\,\frac{5}{2}}}\,\ud x\,, \ldots
\end{align}
\end{subequations}
and so on \emph{ad infinitum}. Technically, all other conservation laws can be obtained recursively in a similar way. Here we provide the recurrent formula for the densities $\set*{\sigma_{\,j}}_{\,j\,=\,-1}^{\,+\,\infty}$ that can be used for this purpose:
\begin{multline}\label{eq:rec}
  \partial^{\,-\,1}\,\sigma_{\,j,\,t}\ =\ \delta_{\,j,\,-\,3}\ -\ \frac{1}{2}\;\biggl[\,g\,(\,v\,)\,\sigma_{\,j}\ +\ g_{\,x\,x}\,(\,v\,)\,\sigma_{\,j}\ +\ g\,(\,v\,)\,\sigma_{\,j,\,x\,x}\ + \\ 
  g\,(\,v\,)\,\sum_{k\,=\,-\,1}^{+\,\infty}\,\sigma_{\,j\,-\,k}\,\sigma_{\,k,\,x}\ +\ g\,(\,v\,)\,\sum_{k\,=\,-\,1}^{+\,\infty}\,\sum_{n\,=\,-\,1}^{+\,\infty}\,\sigma_{\,j\,-\,k\,-\,n}\,\bigl(\,\sigma_{\,k}\,\sigma_{\,n,\,x}\ +\ 2\,\sigma_{\,k,\,x}\,\sigma_{\,n}\,\bigr)\ +\\ 
  2\,g_{\,x}\,(\,v\,)\,\Bigl(\,\sigma_{\,j,\,x}\
  +\ \sum_{k\,=\,-\,1}^{+\,\infty}\,\sigma_{\,j\,-\,k,\,x}\,\sigma_{\,k}\,\Bigr)\,\biggr]\,,
\end{multline}
where $\delta_{\,i\,j}$ is the classical \nm{Kronecker} symbol\footnote{The \nm{Kronecker} symbol $\delta$ is traditionally defined as
\begin{equation*}
  \delta_{\,i\,j}\ =\ \begin{dcases}
   \ 1\,, & \quad i\ =\ j\,, \\
   \ 0\,, & \quad \text{otherwise}.
  \end{dcases}
\end{equation*}}, $\partial^{\,-\,1}$ is a primitive\footnote{Choose $x_{\,0}\ \in\ \R\,$, then
\begin{equation*}
  \partial^{\,-\,1}\,(\,-\,)\ \eqdef\ \int_{\,x_{\,0}}^{\,x}\,(\,-\,)\,\ud x\,.
\end{equation*}
}, $g\,(\,v\,)\ \eqdef\ \dfrac{1}{(\,v\ -\ v_{\,x\,x}\,)^{\,\frac{3}{2}}}$ and it is implicitly assumed above that $\sigma_{\,j}\ \equiv\ \fO\,$, $\forall\,j\ <\ -\,1$ so that all the sums are finite. The recurrent formula \eqref{eq:rec} should be first used with $j\ =\ -\,3$ to determine $\sigma_{\,-\,1}\,$. Then, we fix $j\ =\ -\,2$ and find $\sigma_{\,0}$ and so on \emph{ad infinitum}.

All these conservation laws generate naturally an infinite hierarchy of mutually commuting \nm{Hamiltonian} flows with respect to evolution parameters $t_{\,j}\ \in\ \R\,$:
\begin{equation*}
  v_{\,t_{\,j}}\ =\ -\,\vartheta\,\grad_{\,v}\,\gamma_{\,j}\,[\,v\,]\ \defeq\ K_{\,j}\,[\,v\,]\,, \qquad \forall\,j\ \in\ \Z^{\,+}\,\cup\,\set*{-\,1}\,.
\end{equation*}
Moreover, we have that
\begin{equation*}
  \lb{K_{\,j}}{K_{\,s}}\ =\ \fO\,, \qquad \pb{\gamma_{\,j}}{\gamma_{\,s}}\ =\ \fO\,, \qquad \forall\,j,\,s\ \in\ \Z^{\,+}\,\cup\,\set*{-\,1}\,.
\end{equation*}
Thus, we can deduce our dynamical system \eqref{eq:2.5} and, respectively, its commuting symmetry \eqref{eq:2.4} are jointly integrable \nm{Hamiltonian} flows on the functional manifold $\M\,$. Taking into account the resulting existence of the recursion operator $\Lambda$ \eqref{eq:lambda}, satisfying the second equation \eqref{eq:2.5} along with associated gradient recursion relationships
\begin{equation}\label{eq:2.17}
  \Lambda\,\grad\,\gamma_{\,j}\ =\ \grad\,\gamma_{\,j\,+\,2}\,, \qquad \forall\,j\ \in\ \Z^{\,+}\,\cup\,\set*{-\,1}\,,
\end{equation}
we also deduce the existence of the second \nm{Poisson} operator
\begin{equation*}
  \eta\ \eqdef\ \vartheta\,\Lambda\,:\ \T^{\,\ast}\,(\,\M\,)\ \longrightarrow\ \T\,(\,\M\,)\,,
\end{equation*}
that is compatible with the first $\vartheta$ on the functional manifold $\M\,$. In other terms, the sum
\begin{equation*}
  (\,\vartheta\ +\ \lambda\,\eta\,)\,:\ \T^{\,\ast}\,(\,\M\,)\ \longrightarrow\ \T\,(\,\M\,)
\end{equation*}
persists to be \nm{Poissonian} $\forall\,\lambda\ \in\ \R\,$. The exact expression for the second \nm{Poisson} operator can be easily obtained from the operator relationship, following naturally from the \nm{Lax--Noether} equation \eqref{eq:2.9} in the case when its solution $\psi\ \in\ \T^{\,\ast}\,(\,\M\,)$ is not symmetric, \ie $\psi^{\,\prime}\ \neq\ \psi^{\,\prime\,\ast}$ on the whole manifold $\M\,$.

It is easy to check that the conservation law $\H_{\,1}$ defined in \eqref{eq:2.4b} can be represented in the following equivalent form
\begin{equation*}
  \H_{\,1}\ =\ \bigl(\,\third\;v\,v_{\,x}\ -\ \third\;\partial^{\,-\,1}\,v^{\,2}\,\vert\,v_{\,x}\,\bigr)\ +\ \bigl(\,\partial^{\,-\,1}\,v\ -\ v_{\,x}\,\vert\,v_{\,x}\,\bigr)\ \equiv\ (\,\psi_{\,1}\,\vert\,v_{\,x}\,)
\end{equation*}
with respect to the standard bi-linear form \eqref{eq:bil}, where
\begin{equation*}
  \psi_{\,1}\ \eqdef\ \third\;v\,v_{\,x}\ -\ \third\;\partial^{\,-\,1}\,v^{\,2}\ +\ \partial^{\,-\,1}\,v\ -\ v_{\,x}\ \in\ \T^{\,\ast}\,(\,\M\,)
\end{equation*}
does satisfy the \nm{Lax--Noether} equation \eqref{eq:2.9}
\begin{equation*}
  \psi_{\,t}\ +\ \K^{\,\prime\,\ast}\,\psi\ =\ \fO \pmod{\grad\,L}
\end{equation*}
for some smooth functional $L\,:\ \M\ \longrightarrow\ \R$ on the manifold $\M\,$. Then, it is easy to observe that the operator
\begin{multline*}
  \eta_{\,1}^{\,-\,1}\ =\ \psi_{\,1}^{\,\prime}\ -\ \psi_{\,1}^{\,\prime\,\ast}\ =\ 2\,(\,\partial^{\,-\,1}\ -\ \partial\,)\ +\ \third\;(\,\partial\,v\ +\ v\,\partial\,)\ -\ \twothirds\;(\,v\,\partial^{\,-\,1}\ +\ \partial^{\,-\,1}\,v\,)\ =\\ 
  2\,\vartheta^{\,-\,1}\ +\ \twothirds\;\bigl(\,\sqrt{v}\,\partial\,\sqrt{v}\ -\ 2\,\partial^{\,-\,1}\,\sqrt{v}\,\partial^{\,-\,1}\,\bigr)\,.
\end{multline*}
is \emph{a priori} symplectic on the functional manifold $\M$ jointly with the expression
\begin{equation}\label{eq:2.21}
  \eta_{\,2}^{\,-\,1}\ \eqdef\ \sqrt{v}\,\partial\,\sqrt{v}\ -\ 2\,\partial^{\,-\,1}\,\sqrt{v}\,\partial\,\sqrt{v}\,\partial^{\,-\,1}\,,
\end{equation}
since a linear combination of symplectic non-degenerate operators is always symplectic if it is non-degenerate \cite{Abraham1987, Arnold1997}. Similarly, one can find that
\begin{equation*}
  \H_{\,2}\ =\ 2\,\int_{\,T}\sqrt{v\ -\ v_{\,x\,x}}\,\ud x\ =\ \biggl(\,(\,\partial\ -\ \partial^{\,-\,1}\,)\,\Bigl(\,\frac{2}{\sqrt{v\ -\ v_{\,x\,x}}}\,\Bigr)\,\Big\vert\,v_{\,x}\,\biggr)\ \equiv\ (\,\psi_{\,2}\,\vert\,v_{\,x}\,)\,,
\end{equation*}
provided that the generating element $\psi_{\,2}\ \eqdef\ (\,\partial\ -\ \partial^{\,-\,1}\,)\,\Bigl(\,\frac{2}{\sqrt{v\ -\ v_{\,x\,x}}}\,\Bigr)\ \in\ \T^{\,\ast}\,(\,\M\,)$ which gives rise to the second compatible symplectic operator
\begin{equation*}
  \eta_{\,3}^{\,-\,1}\ \eqdef\ \psi_{\,2}^{\,\prime}\ -\ \psi_{\,2}^{\,\prime\,\ast}\ =\ (\,\partial^{\,-\,1}\ -\ \partial\,)\,\biggl(\,\frac{1}{(\,v\ -\ v_{\,x\,x}\,)^{\,\frac{3}{2}}}\;\partial\ +\ \partial\;\frac{1}{(\,v\ -\ v_{\,x\,x}\,)^{\,\frac{3}{2}}}\,\biggr)\,(\,\partial^{\,-\,1}\ -\ \partial\,)\,.
\end{equation*}
Moreover, as there holds the representation $\eta_{\,3}^{\,-\,1}\ \equiv\ \vartheta^{\,-\,1}\,\eta\,\vartheta^{\,-\,1}\,$, we can easily deduce that the operator expression
\begin{equation}\label{eq:2.24}
  \eta\ \eqdef\ \frac{1}{(\,v\ -\ v_{\,x\,x}\,)^{\,\frac{3}{2}}}\;\partial\ +\ \partial\;\frac{1}{(\,v\ -\ v_{\,x\,x}\,)^{\,\frac{3}{2}}}
\end{equation}
provides us the third compatible \nm{Poisson} operator $\eta\,:\ \T^{\,\ast}\,(\,\M\,)\ \longrightarrow\ \T\,(\,\M\,)$ for the \nm{Hamiltonian} system \eqref{eq:2.5} on the functional manifold $\M$ and determines the recursion operator
\begin{equation*}
  \Lambda\ =\ \eta^{\,-\,1}\,\vartheta\ \equiv\ \bigl(\,\sqrt{v}\,\partial\,\sqrt{v}\ -\ 2\,\partial^{\,-\,1}\,\sqrt{v}\,\partial\,\sqrt{v}\,\partial^{\,-\,1}\,\bigr)\,(\,\partial^{\,-\,1}\ -\ \partial\,)^{\,-\,1}\,,
\end{equation*}
that satisfies the relationships \eqref{eq:2.17}. The recursive scheme can be naturally extended to the negative values of indices as well:
\begin{equation*}
  \Lambda\,\grad\,\gamma_{\,j}\ =\ \grad\,\gamma_{\,j\,+\,2}\,, \qquad j\ \in\ \Z\,.
\end{equation*}
In particular, one has
\begin{equation*}
  \Lambda^{\,-\,1}\,\grad\,\gamma_{\,-\,1}\ =\ \grad\,\gamma_{\,-\,3}\,,
\end{equation*}
with $\gamma_{\,-\,3}\ \equiv\ \H_{\,1}\,$. This observation yields right away that the \nm{Hamiltonian} flow \eqref{eq:2.4} possesses an infinite hierarchy of commuting to each other conservation laws \eqref{eq:2.14}. Thus, it is a completely integrable \nm{Hamiltonian} system on the functional manifold $\M$ \cite{Mitropolsky1987, Baszak1998}. Finally, we can summarize the main finding of the present study in the following
\begin{theorem}
The nonlinear dynamical system \eqref{eq:2.5} jointly with its symmetry\footnote{To be more precise, in \cite{Cheviakov2021} it was shown that \eqref{eq:2.4} possesses a \nm{Lie--B\"acklund} symmetry given by the flow \eqref{eq:2.5}. However, this relation is \emph{symmetric} since
\begin{equation*}
  \lb*{\tilde{\Kk}}{\tilde{\F}}\ =\ \fO\ \qquad \Longrightarrow\qquad\ \lb*{\tilde{\F}}{\tilde{\Kk}}\ =\ \fO\,.
\end{equation*}} \eqref{eq:2.4} possess a common infinite hierarchy of mutually commuting conservation laws \eqref{eq:2.14} with respect to three compatible \nm{Poisson} brackets \eqref{eq:1.6}, \eqref{eq:2.21} and \eqref{eq:2.24}, and are completely integrable \nm{Hamiltonian} systems on the smooth functional \nm{Poisson} manifold $\M\,$.
\end{theorem}


\section{Solitary wave solutions and numerical illustrations}
\label{sec:num}

In this Section, we would like to study the solitary wave solutions to \cref{eq:abbm} numerically in the proven above integrable (eBBM${}_{\,\third}$ \cref{eq:bbm3}) and non-integrable ($\alpha\ \neq\ \frac{1}{3}$) cases.

\begin{figure}
  \centering
  \subfigure[$c\ =\ 1.2$]{\includegraphics[width=0.48\textwidth]{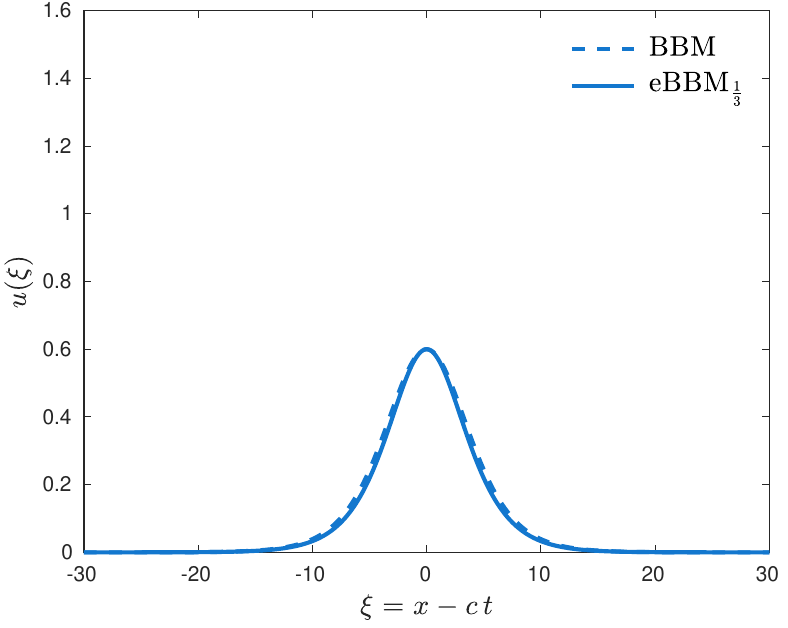}}
  \subfigure[$c\ =\ 1.3$]{\includegraphics[width=0.48\textwidth]{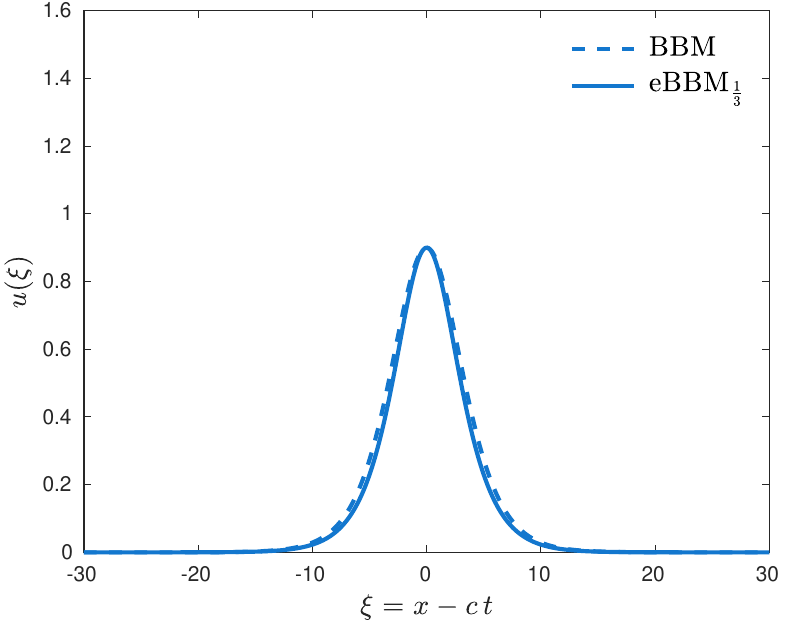}}
  \subfigure[$c\ =\ 1.4$]{\includegraphics[width=0.48\textwidth]{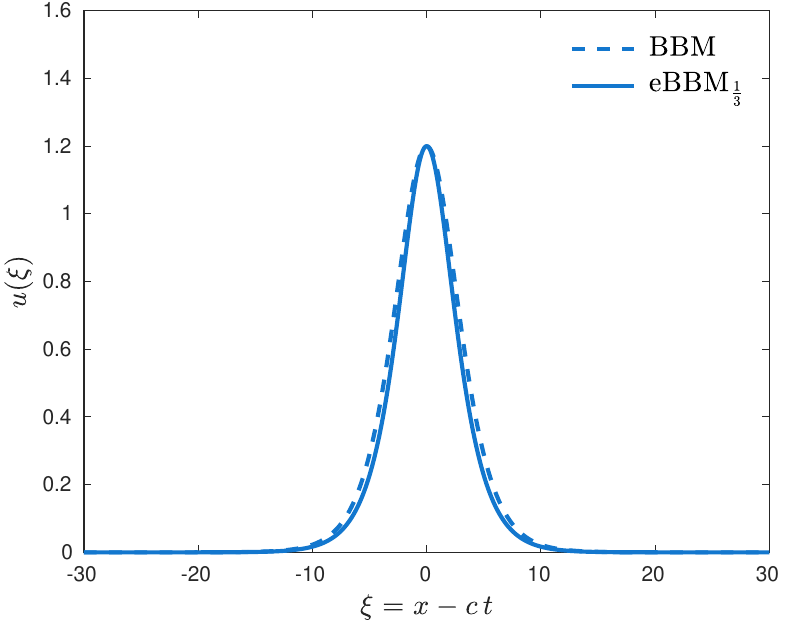}}
  \subfigure[$c\ =\ 1.5$]{\includegraphics[width=0.48\textwidth]{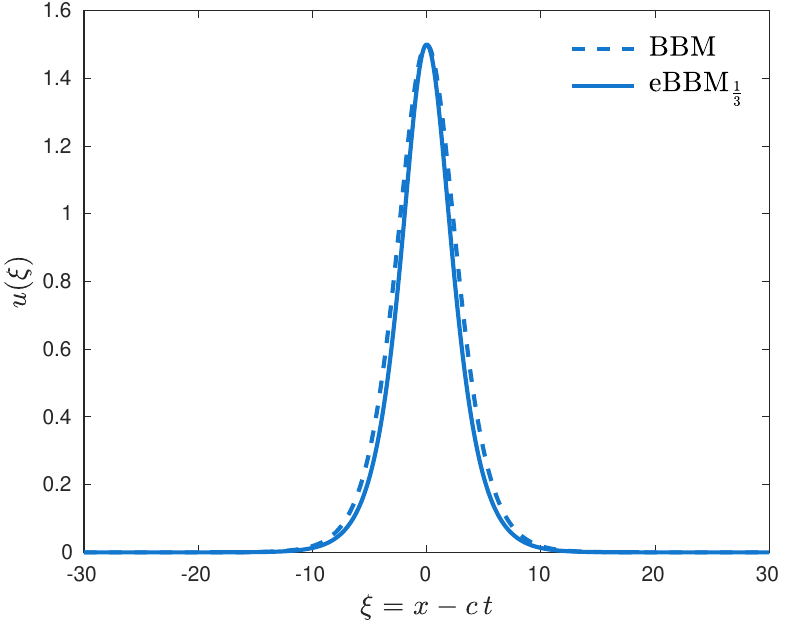}}
  \caption{\small\em Solitary wave solutions to the classical BBM \eqref{eq:bbm} and integrable eBBM${}_{\,\third}$ equation \eqref{eq:bbm3}.}
  \label{fig:sw}
\end{figure}

First of all, let us remind the reader that the explicit solitary wave solutions to the non-integrable BBM equation (the particular case of \cref{eq:abbm} with $\alpha\ =\ 0$) are well known:
\begin{equation}\label{eq:bbmsw}
  u\,(\,x,\,t\,)\ =\ 3\,(\,c\ -\ 1\,)\,\sech^{\,2}\,\biggl(\,\frac{1}{2}\;\sqrt{\frac{c\ -\ 1}{c}}\;(\,x\ -\ c\,t\,)\,\biggr)\,,
\end{equation}
where $c\ \in\ \R$ is the solitary wave travelling velocity. Unfortunately, no other solutions to \cref{eq:abbm} of this (analytical) form exist for any $\alpha\ \neq\ 0\,$. Fortunately, there are other analytical and numerical methods available. Using the classical \nm{Petviashvili} iterative method \cite{Petviashvili1976}, we computed numerically the solitary wave solutions to the integrable eBBM${}_{\,\third}$ \cref{eq:bbm3}. In Figure~\ref{fig:sw}, we compare the corresponding solitary wave shapes with analytical solutions \eqref{eq:bbmsw} to the BBM equation for several values of the propagation speed parameter $c\,$. From these comparisons, one can conclude that the BBM equation has slightly thicker solitary waves for the same fixed amplitude. Moreover, all these solitary waves share the same speed/amplitude relation. The last fact is not difficult to demonstrate analytically.

\begin{prop}\label{prop:1}
Localized\footnote{By localization we mean that the function $u\,(\,\xi\,)$ verifies the following boundary conditions at infinity:
\begin{equation*}
  \lim_{\xi\,\to\,\infty}\,u^{\,(\,n\,)}\,(\,\xi\,)\ =\ 0\,,
\end{equation*}
where $u^{\,(\,n\,)}$ denotes the $n$\up{th} derivative of the smooth function $u\,$, \ie $u^{\,(\,1\,)}\ \equiv\ u^{\,\prime}\,$, $u^{\,(\,2\,)}\ \equiv\ u^{\,\dprime}\,$, \etc} travelling solutions $u\,(\,x,\,t\,)\ \coloneq\ u\,(\,\xi\,)$ with $\xi\ \eqdef\ x\ -\ c\,t\ \in\ \R$ verify the following Ordinary Differential Equation:
\begin{equation}\label{eq:ode}
  \frac{1\ -\ c}{2}\;u^{\,2}\ +\ \frac{1}{6}\;u^{\,3}\ +\ \frac{c}{2}\;(\,u^{\,\prime}\,)^{\,2}\ -\ \frac{\alpha}{2}\;u\,(\,u^{\,\prime}\,)^{\,2}\ =\ \fO\,,
\end{equation}
where the prime denotes throughout this Section the usual derivative with respect to the variable $\xi\,$.
\end{prop}

\begin{proof}
The proof is done by substituting the travelling wave ansatz into \cref{eq:abbm} and performing two successive integrations with the first integrating factor $1$ and $u^{\,\prime}$ right afterwards.
\end{proof}

\begin{corollary}\label{cor:1}
The speed--amplitude relation $a\ \mapsto\ c$ for solitary wave solutions to the $\alpha-$family of PDEs \eqref{eq:abbm} is independent of $\alpha$ and is explicitly given by
\begin{equation}\label{eq:sa}
  a\ =\ 3\cdot(\,c\ -\ 1\,)\,.
\end{equation}
\end{corollary}

\begin{proof}
To fix the ideas, let us assume that the crest is located at $\xi\ =\ 0\,$. At the solitary wave crest, we have, by definition:
\begin{equation*}
  u\,(\,0\,)\ =\ a\ \geq\ 0\,, \qquad u^{\,\prime}\,(\,0\,)\ =\ 0\,.
\end{equation*}
By taking the limit $\xi\ \to\ 0$ in \eqref{eq:ode} and, after substituting the above values, we obtain the desired result \eqref{eq:sa}.
\end{proof}
It follows obviously from \eqref{eq:sa} that the solitary wave speed $c$ must be supra-linear ($c\ >\ 1$) to have solitary waves of positive amplitude. The solitary waves of depression are out of the scope of the present study because the $\alpha-$family of PDEs \eqref{eq:abbm} is considered as a model for long gravity free surface waves. The (physical) solitary waves of depression do not exist in the $\alpha-$family \eqref{eq:abbm}.

\cref{cor:1} sheds some light on the numerical results shown in \cref{fig:sw}. Namely, the speed-amplitude relation will be the same even if we compute the solitary wave solutions to any other value of $\alpha$ different from $\set*{0,\,\third}\,$. It turns out that \cref{eq:ode} can be used to derive approximate solitary waves to \eqref{eq:abbm} for small values of the parameter $\alpha\,$:
\begin{equation}\label{eq:ass}
  u\,(\,\xi\,)\ =\ \sum_{j\,=\,0}^{N}\,u_{\,j}\,(\,\xi\,)\,\alpha^{\,j}\ +\ \O\,(\,\alpha^{\,N\,+\,1}\,)\,,
\end{equation}
where $u_{\,0}\,(\,\xi\,)$ is given in \eqref{eq:bbmsw}:
\begin{equation*}
  u_{\,0}\,(\,\xi\,)\ =\ 3\,(\,c\ -\ 1\,)\,\sech^{\,2}\,\biggl(\,\frac{1}{2}\;\sqrt{\frac{c\ -\ 1}{c}}\;\xi\,\biggr)\,.
\end{equation*}
We also provide below a few higher-order terms in the asymptotic expansion \eqref{eq:ass}:
\begin{equation*}
  u_{\,1}\,(\,\xi\,)\ =\ \frac{9\,(\,c\ -\ 1\,)^{\,2}}{c}\;\frac{\sinh\,(\,\cosh\ -\ \sinh\,)}{\cosh^{\,4}}\,,
\end{equation*}
\begin{equation*}
  u_{\,2}\,(\,\xi\,)\ =\ \frac{9\,(\,c\ -\ 1\,)^{\,3}}{2\,c^{\,2}}\;\frac{10\,\cosh^{\,4}\ -\ 10\,\cosh^{\,3}\,\sinh\ +\ 24\,\cosh\,\sinh\ -\ 29\,\cosh^{\,2}\ +\ 16}{(\,\cosh2\ -\ 1\,)\,\cosh^{\,6}}\,,
\end{equation*}
where, for the sake of brevity, we defined:
\begin{align*}
  \cosh\ &\eqdef\ \cosh\,\biggl(\,\frac{1}{2}\;\sqrt{\frac{c\ -\ 1}{c}}\;\xi\,\biggr)\,, \\
  \sinh\ &\eqdef\ \sinh\,\biggl(\,\frac{1}{2}\;\sqrt{\frac{c\ -\ 1}{c}}\;\xi\,\biggr)\,, \\
  \cosh2\ &\eqdef\ \cosh\,\biggl(\,\sqrt{\frac{c\ -\ 1}{c}}\;\xi\,\biggr)\,.
\end{align*}

\begin{figure}
  \centering
  \includegraphics[width=0.99\textwidth]{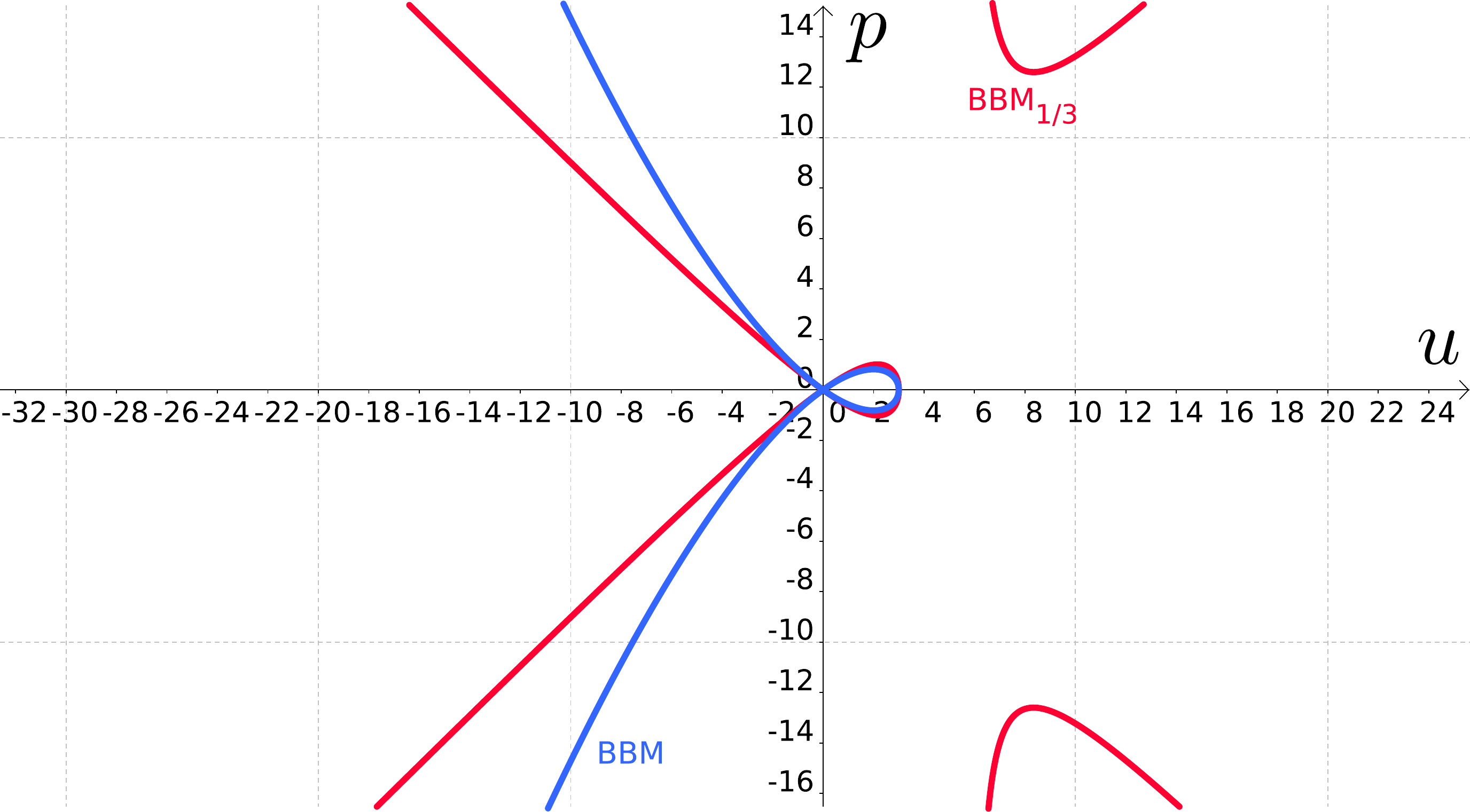}
  \caption{\small\em Implicit curves in the phase plane $(\,u,\,p\ \equiv\ u^{\,\prime}\,)$ defined by the algebraic equation~\eqref{eq:alg} for $c\ =\ 2\,$, $\alpha\ =\ 0$ (blue curve) and $\alpha\ =\ \frac{1}{3}$ (red curve).}
  \label{fig:phase}
\end{figure}

From \cref{prop:1} we can also deduce that \cite{Dutykh2016, Dutykh2019}
\begin{corollary}\label{cor:2}
Solitary wave solution to \cref{eq:abbm} belong to the following family of algebraic curves defined on the phase plane $(\,u,\,u^{\,\prime}\,)\ \defeq\ (\,u,\,p\,)\ \in\ \R^{\,2}\,$:
\begin{equation}\label{eq:alg}
  \Q_{\,\alpha,\,c}\ \eqdef\ \Set*{(\,u,\,p\,)\ \in\ \R^{\,2}}{\frac{1\ -\ c}{2}\;u^{\,2}\ +\ \frac{1}{6}\;u^{\,3}\ +\ \frac{c}{2}\;p^{\,2}\ -\ \frac{\alpha}{2}\;u\,p^{\,2}\ =\ \fO}\,,
\end{equation}
\end{corollary}
For illustration, we show in \cref{fig:phase} two algebraic curves defined in \cref{eq:alg}. We underline that physical solitary waves are supported by a small loop situated on the right to the origin. Other branches do not correspond to physical solutions.

\subsection{Solitary waves interactions}

\begin{figure}
  \centering
  \includegraphics[width=0.99\textwidth]{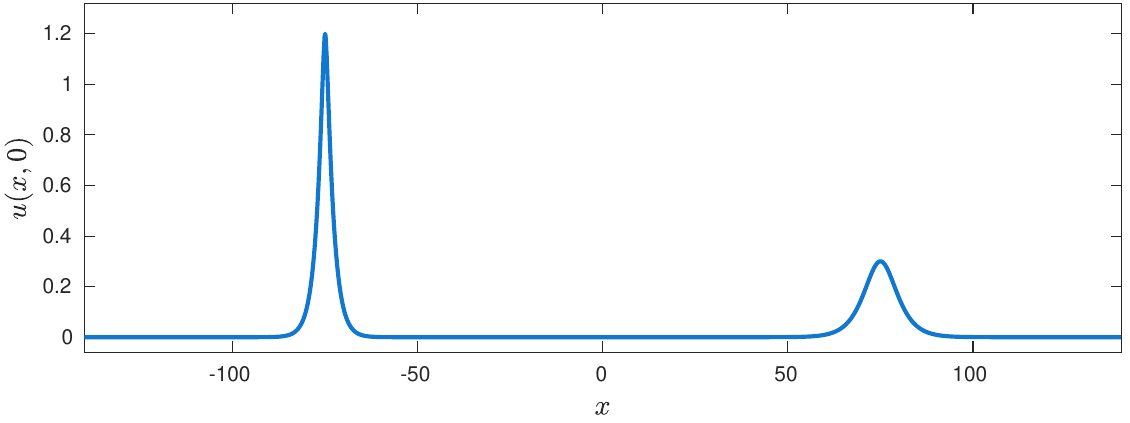}
  \caption{\small\em The initial condition used for the overtaking solitary wave collision experiments.}
  \label{fig:ic}
\end{figure}

\begin{figure}
  \centering
  \subfigure[BBM equation \eqref{eq:bbm}]{\includegraphics[width=0.99\textwidth]{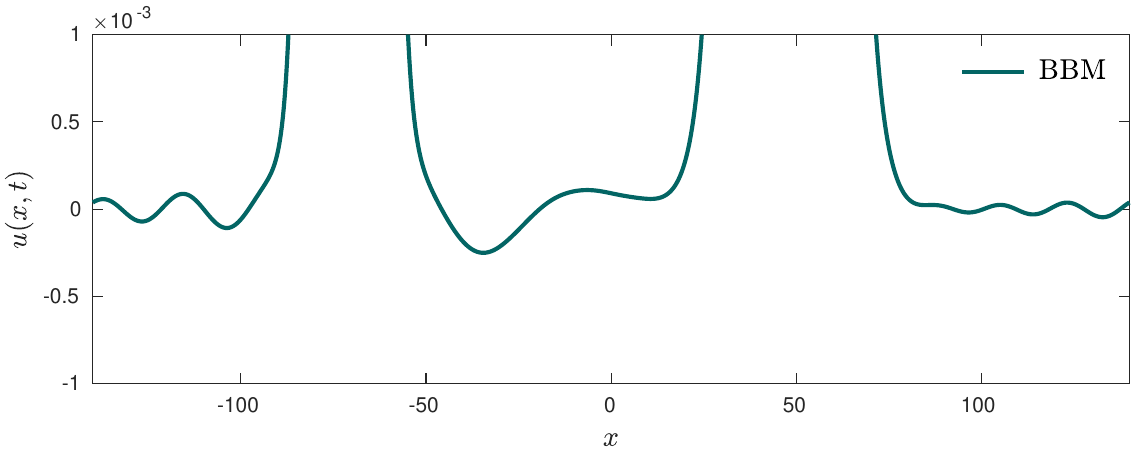}}
  \subfigure[eBBM equation \eqref{eq:ebbm}]{\includegraphics[width=0.99\textwidth]{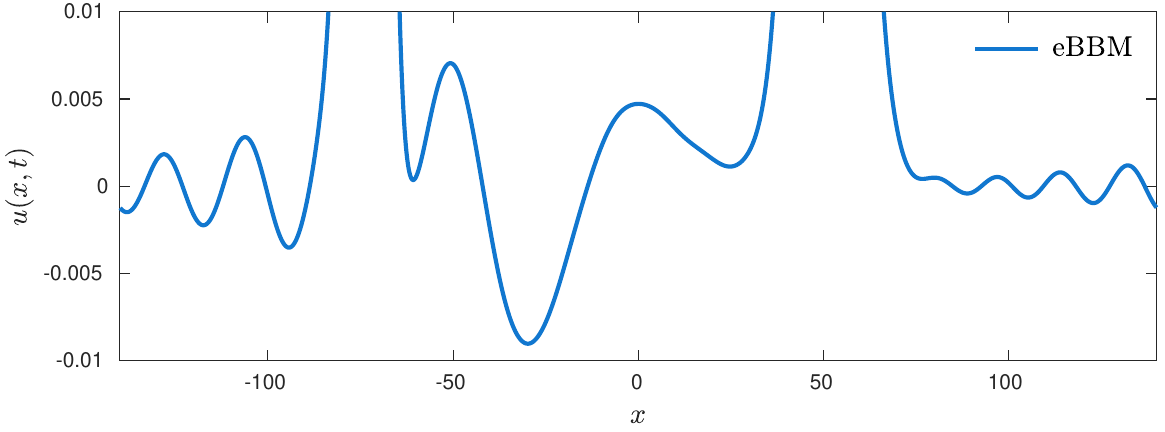}}
  \subfigure[eBBM${}_{\,\third}$ equation \eqref{eq:bbm3}]{\includegraphics[width=0.99\textwidth]{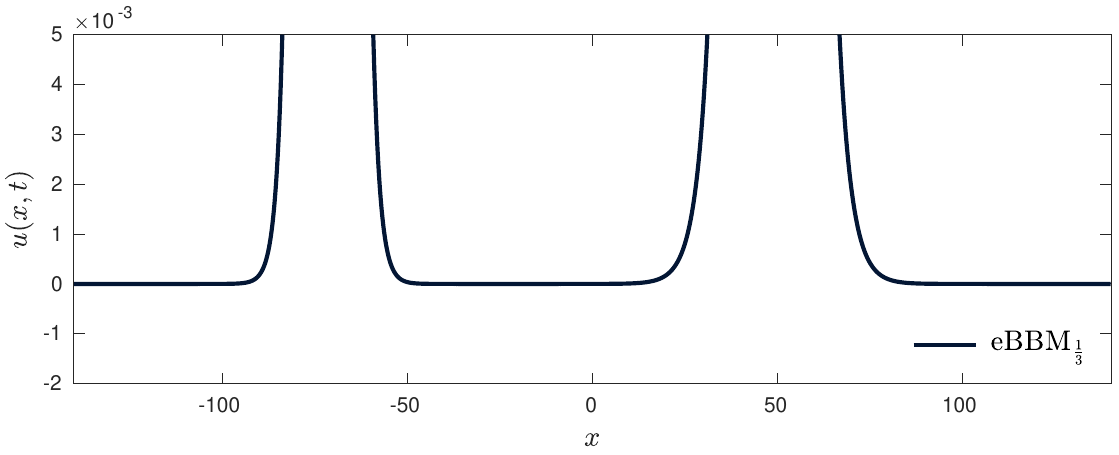}}
  \caption{\small\em Zoom on the overtaking collision of two solitary waves at the final simulation time $T\ =\ 1000$ in three different BBM-type equations. Please notice the difference in the vertical scales.}
  \label{fig:coll}
\end{figure}

As a final test case, we consider the interaction of two solitary waves under the dynamics of the $\alpha-$family of PDEs \eqref{eq:abbm}. Namely, we shall consider three cases:
\begin{itemize}
  \item BBM equation \eqref{eq:bbm} ($\alpha\ =\ 0$),
  \item eBBM equation\footnote{This equation is \nm{Galilean} invariant and energy preserving \cite{Cheviakov2023}.} equation \eqref{eq:ebbm} ($\alpha\ =\ 1$).
  \item eBBM${}_{\,\third}$ equation \eqref{eq:bbm3} ($\alpha\ =\ \third$),
\end{itemize}
We consider the same initial value problem, which will be simulated using a \nm{Fourier}-type pseudo-spectral method \cite{Dutykh2011a} in all three partial differential equations mentioned above. Namely, we consider a computational domain given by the 1D torus $[\,-\,140,\,140\,]\ \subset\ \R\,$. Two right-propagating solitary waves are initially located at positions $x_{\,0}\ =\ \pm\,75\,$. The left solitary wave has the propagation velocity\footnote{The corresponding solitary wave amplitudes can be computed according to the speed--amplitude relation \eqref{eq:sa}.} $c_{\,1}\ =\ 1.4$ and the right --- $c_{\,2}\ =\ 1.1\,$. This initial condition is shown in \cref{fig:ic}. Then, we let this initial condition evolve from $t\ =\ 0$ to $t\ =\ T\ \coloneq\ 1000\,$. In our numerical simulations, we used $N\ =\ 16\,384$ \nm{Fourier} modes together with the classical $2/3$\up{rd} anti-aliasing rule. We used the adaptive \texttt{ode113} solver \cite{Shampine1997} for the time integration. During this interval of time, the two solitary waves have the time to interact in the overtaking collision. The results of this interaction are shown in Figure \ref{fig:coll}:
\begin{description}
  \item[Panel (\textit{a})] BBM equation \eqref{eq:bbm},
  \item[Panel (\textit{b})] eBBM equation \eqref{eq:ebbm},
  \item[Panel (\textit{c})] eBBM${}_{\,\third}$ equation \eqref{eq:bbm3}.
\end{description}
In perfect agreement with our theoretical investigations, the solitary wave interaction is elastic exclusively in the integrable case \eqref{eq:bbm3}. The interaction process produces small radiation ($\alpha\ \neq\ \third$). The magnitude of observed oscillations may be measured for the given equation `\emph{non-integrability}'\footnote{In the sense of the deviation from the integrability property.}.

\section{Conclusions and perspectives}
\label{sec:concl}

The present study is rooted in modelling nonlinear and dispersive shallow water waves combined with the \nm{Lie} symmetry analysis. It has been noticed in \cite{Cheviakov2021} that the eBBM${}_{\,\third}$ \cref{eq:bbm3} is nothing else but a rescaled version of the celebrated \nm{Camassa--Holm} equation \cite{Johnson2002, Dullin2003}. This observation implies the integrability of \eqref{eq:bbm3}. However, in the present work, we provide an alternative direct mathematical proof of the same fact with a novel mathematical method. Moreover, as a byproduct of our analysis, we establish the integrability of another dynamical system. Namely, Equation~\eqref{eq:2.5} is a \nm{Lie--B\"acklund} symmetry of the rescaled eBBM${}_{\,\third}$ Equation~\eqref{eq:2.4}. Based on the gradient-holonomic integrability checking scheme applied to these dynamical systems, we analytically constructed its associated infinite hierarchy of conservation laws, derived three compatible \nm{Poisson} structures, and established its complete integrability. The present study is another demonstration of the gradient-holonomic scheme. In future works, we plan to apply this scheme to other nonlinear dynamical systems arising in various application domains. Finally, this study was completed by an analytical and numerical investigation of the travelling wave solutions of the $\alpha-$family \eqref{eq:abbm} in terms of their shapes and interactions on the real line.


\subsection*{Acknowledgments}
\addcontentsline{toc}{subsection}{Acknowledgments}

Authors are thankful to Prof.~Anatoli K. \nm{Prykarpatski} (Krak\'{o}w University of Technology, Poland) and Prof.~Jan \nm{Koro\'{n}ski} (Krak\'{o}w Polytechnical University, Poland) for valuable discussions, comments and remarks on the manuscript.


\subsection*{Funding}
\addcontentsline{toc}{subsection}{Funding}

This publication is based upon work supported by the Khalifa University of Science and Technology under Award No. FSU$-2023-014$.


\subsection*{Author contributions}
\addcontentsline{toc}{subsection}{Author contributions}

All the Authors contributed equally to this work.


\subsection*{Conflict of interest statement}
\addcontentsline{toc}{subsection}{Conflict of interest statement}

The Authors certify that they have \textbf{no} affiliations with or involvement in any organization or entity with any financial interest (such as honoraria; educational grants; participation in speakers' bureaus; membership, employment, consultancies, stock ownership, or other equity interest; and expert testimony or patent-licensing arrangements), or non-financial interest (such as personal or professional relationships, affiliations, knowledge or beliefs) in the subject matter or materials discussed in this manuscript.


\subsection*{Disclaimer statement}
\addcontentsline{toc}{subsection}{Disclaimer statement}

The opinions expressed in this publication/study are those of the Authors and do not necessarily reflect the views of their employers or any other affiliated organizations.



\bigskip\bigskip
\invisiblesection{References}
\bibliographystyle{acm}
\bibliography{mybiblio}
\bigskip\bigskip


\TheEnd

\end{document}